\documentclass[%
 reprint,revtex4-2 Template and Sample (1)
preprint,
 amsmath,amssymb,
 aps,
 prl,
twocolumn
]{revtex4-2}
\usepackage[version=3]{mhchem}

\usepackage{multirow}
\usepackage{booktabs}
\usepackage{amsmath}
\usepackage{amssymb}
\usepackage{graphicx}
\usepackage{dcolumn}
\usepackage{bm}
\usepackage[mathlines]{lineno}
\usepackage{xcolor}
\usepackage{comment}
\usepackage[unicode=true,pdfusetitle,bookmarks=false,colorlinks=true,citecolor=blue,urlcolor=blue,linkcolor=blue]{hyperref}

\usepackage{algorithm}
\usepackage[noend]{algpseudocode}
\usepackage{etoolbox}

\makeatletter
\newcommand*{\algrule}[1][\algorithmicindent]{%
  \makebox[#1][l]{%
    \hspace*{.2em}% <------------- This is where the rule starts from
    \vrule height .75\baselineskip depth .25\baselineskip
  }
}

\newcount\ALG@printindent@tempcnta
\def\ALG@printindent{%
    \ifnum \theALG@nested>0% is there anything to print
    \ifx\ALG@text\ALG@x@notext% is this an end group without any text?
    \else
    \unskip
    \ALG@printindent@tempcnta=1
    \loop
    \algrule[\csname ALG@ind@\the\ALG@printindent@tempcnta\endcsname]%
    \advance \ALG@printindent@tempcnta 1
    \ifnum \ALG@printindent@tempcnta<\numexpr\theALG@nested+1\relax
    \repeat
    \fi
    \fi
}
\patchcmd{\ALG@doentity}{\noindent\hskip\ALG@tlm}{\ALG@printindent}{}{\errmessage{failed to patch}}
\patchcmd{\ALG@doentity}{\item[]\nointerlineskip}{}{}{} % no spurious vertical space
\makeatother

\begin{document}

\title{Inferring Surface Slip in Active Colloids from Flow Fields \\ Using Physics-Informed Neural Networks}% Force line breaks with \\

\author{Parvin Bayati}
\affiliation{Department of Chemistry, The Pennsylvania State University, University Park, Pennsylvania, 16802, USA}

\author{Stewart A. Mallory}
\email{sam7808@psu.edu}
\affiliation{Department of Chemistry, The Pennsylvania State University, University Park, Pennsylvania, 16802, USA}
\affiliation{Department of Chemical Engineering, The Pennsylvania State University, University Park, Pennsylvania, 16802, USA}


\begin{abstract}
The directed motion of active colloids is governed by spatial variations in surface chemistry and interfacial stress, yet these properties remain extremely difficult to measure directly.
We introduce a physics-informed neural network framework that infers the slip distribution driving propulsion from partial observations of the surrounding flow.
By combining sparse fluid velocity measurements with the Stokes equations and boundary constraints, the method reconstructs both the near-surface slip and the full velocity and pressure fields.
Validation against analytical solutions and Boundary Element Method calculations for canonical active colloid models shows quantitative agreement in both unbounded and confined geometries.
Crucially, the framework recovers the surface slip even when no flow data are available near the particle, demonstrating that accessible bulk measurements encode the interfacial stresses responsible for active motion.
These results establish physics-informed inference as a powerful tool for characterizing and ultimately controlling interfacially driven transport in colloidal active matter.
\end{abstract}

\maketitle
 
\begin{figure*}[t!]
\centering
\includegraphics[width=0.9\textwidth]{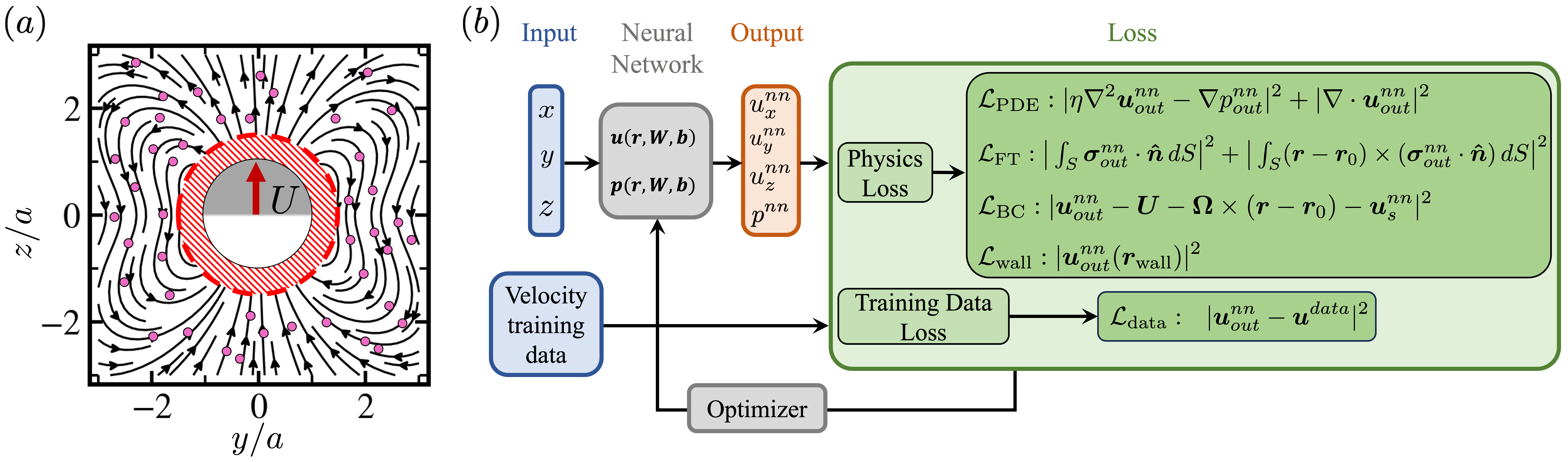}
\caption{\protect\small{
(a) Bulk flow around an active colloid can be measured accurately, but data within the near-surface region (red hatched band) are typically noisy or missing, obscuring the slip velocity and interfacial stresses that generate propulsion.
(b) Spatial coordinates $\boldsymbol r=(x,y,z)$ serve as inputs, and the network outputs $(u_x^{nn},u_y^{nn},u_z^{nn},p^{nn})$.
Physics-based losses enforce the Stokes equations, incompressibility, force- and torque-free conditions, boundary constraints, and no-slip walls when present, while a data-loss term penalizes deviations from measured velocities.
Minimizing the total loss yields physically consistent reconstructions of the flow field and surface slip distribution.
}}
\label{fig:figure_1} 
\end{figure*}

\textit{Introduction.}---Interfacial-driven flows arise from spatial variations in mechanical stress generated by gradients in interfacial tension, temperature, concentration, surface charge, or chemical activity~\cite{scriven1960marangoni,vogel2013comparative,lauga2009hydrodynamics,marchetti2013hydrodynamics}.
Because these stresses mediate momentum transfer between interfaces and the surrounding fluid, they govern how soft materials and complex liquids move, reorganize, and respond to their environments~\cite{drescher2009dancing,li2019interfacial,toyota2009self,weber2019physics}.
A prominent example, and the focus of this work, is colloidal active matter, in which interfacial stresses drive particle self-propulsion.
Spatial variations in surface chemistry or interfacial tension generate surface flows that propel particles through viscous fluids~\cite{paxton2004catalytic,izri2014self,palacci2014light,maass2016swimming,meredith2022chemical}.
These stresses govern propulsion and shape particle–boundary interactions~\cite{suzuno2014maze,shen2018hydrodynamic,zhou2025drag,uspal2015self,bayati2019dynamics,wu2024collective,kawakami2025migration,overberg2024motion,song2022confined,dey2022oscillatory}, particle–particle dynamics~\cite{zottl2014hydrodynamics,bayati2016dynamics,mallory2017self,clopes2022alignment,tucci2024nonreciprocal,liu2024self,nie2023two}, and responses to external fields~\cite{bayati2019electrophoresis,auschra2021thermotaxis,ban2014self,lozon2024chemically,takatori2016acoustic,xiao2022platform,bayati2024orbits}.
More broadly, resolving near-surface flow and slip is essential for understanding the coupling between mechanical transport and biochemical activity, as seen in cortical flows of biological systems such as \textit{Drosophila} embryos, where spatially organized surface contractility coordinates nuclear positioning and cell-cycle oscillations~\cite{deneke2019self,htet2023cortex,htet2025analytical}.
Quantitatively resolving these surface properties is therefore essential for predicting transport and nonequilibrium organization in active matter.

Despite their importance, quantities such as slip velocity, phoretic mobility, and local interfacial stress remain difficult to measure.
Bulk flow fields can be resolved with modern optical imaging techniques~\cite{theunissen2008improvement,kahler2012uncertainty}, but measurements near surfaces are often noisy or incomplete because of optical distortions and tracer–boundary interactions~\cite{katuri2021inferring,singh2022interaction}.
Direct stress measurements are even more limited, since traction force microscopy and interfacial rheometry provide only substrate-restricted or spatially averaged information~\cite{style2014traction,butt2005force}.
As a result, the interfacial properties that govern propulsion often remain experimentally inaccessible.

Inferring these hidden boundary conditions from surrounding flow data requires solving an inverse Stokes problem that is typically ill-posed, highly sensitive to noise, and computationally demanding~\cite{yan2008shape}.
Classical numerical approaches, such as finite and boundary element methods (FEM/BEM), excel at forward simulations, where boundary conditions are prescribed, but are not designed to infer them from partial data.
These limitations motivate frameworks that combine physical constraints with available measurements to extract interfacial information directly from flow fields.

Physics-informed neural networks (PINNs) provide one such route.
By embedding the governing equations into the training loss together with experimental or simulated data, PINNs enforce physical consistency even when measurements are sparse or incomplete~\cite{raissi2019physics,karniadakis2021physics,Cai2021-ee,Zhao2024-sd}.
They have been used to reconstruct flow fields from incomplete observations~\cite{Raissi2020-em,Arzani2021-pg,jin2021nsfnets,gurevich2024learning}, infer pressure from MRI data~\cite{kissas2020machine}, and identify constitutive laws in complex materials~\cite{liu2020generic}.
These prior studies illustrate how physics-constrained learning can be a powerful tool for solving inverse problems that challenge traditional numerical approaches.

Here, we introduce a PINN-based framework that infers the surface slip velocity of an active colloid directly from sparse measurements of the surrounding flow (Fig.~\ref{fig:figure_1}).
The network reconstructs the full velocity and pressure fields by enforcing the Stokes equations and boundary conditions, producing solutions consistent with both the data and the underlying hydrodynamics.
This mesh-free approach is particularly suited to situations where near-surface flow cannot be measured accurately.
By inferring the hidden slip distribution and its associated flow, the method quantitatively links experimentally accessible hydrodynamic fields to the interfacial mechanisms that generate propulsion.

The remainder of the manuscript is organized as follows.
We outline the hydrodynamics of active colloids and introduce the squirmer and phoretic Janus particle models.
We then describe the PINN formulation and validate it using analytical solutions and BEM simulations.
Finally, we discuss the broader implications of this framework and its extensions to more complex systems.

\begin{figure*}[t!]
\centering
\includegraphics[width=0.9\textwidth]{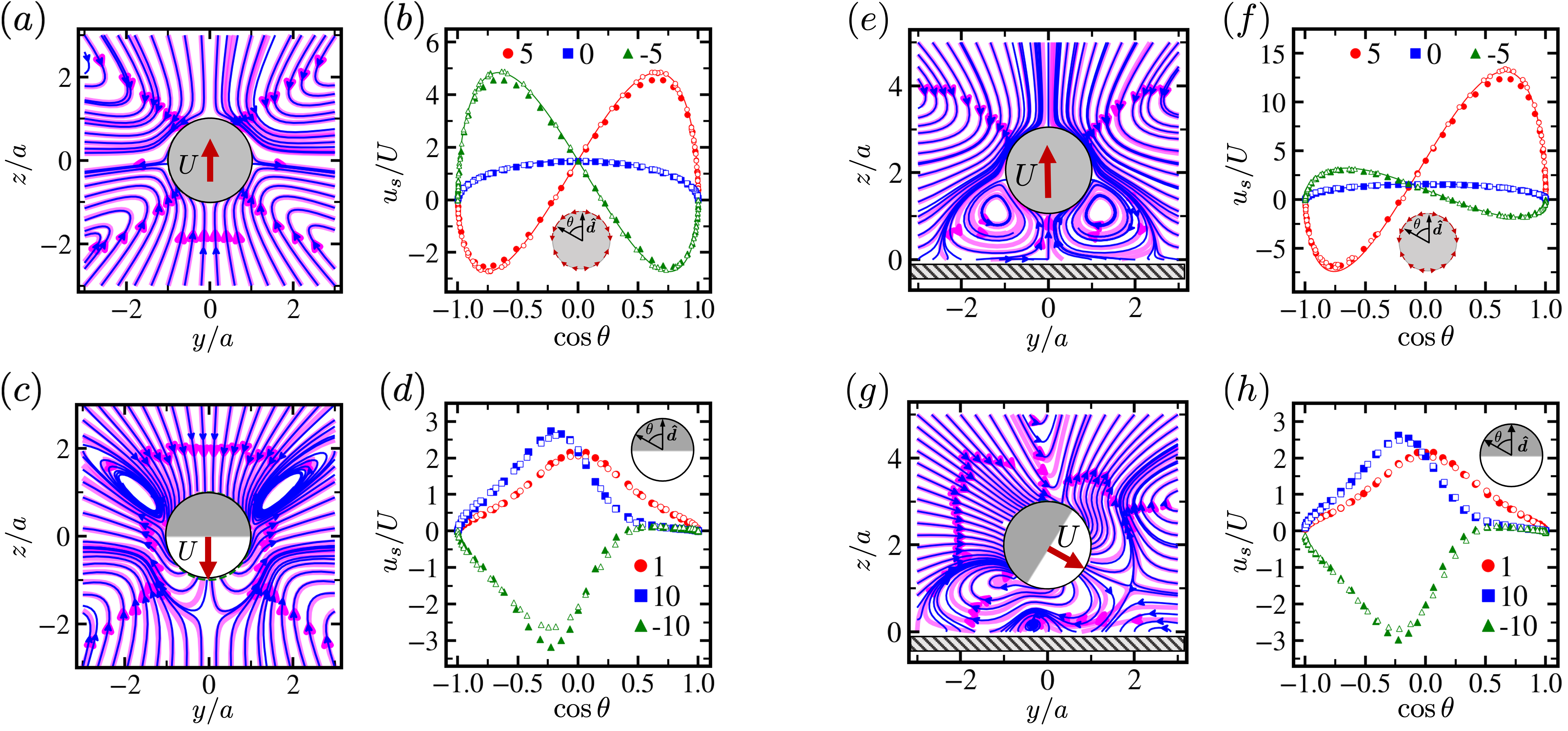}
\caption{\protect\small{
Comparison of PINN-predicted (blue) and BEM (pink) flow fields in the $yz$-plane for squirmer and phoretic Janus particles in the laboratory frame.
Panels (a,c) show unbounded cases: a puller squirmer with $\beta_s = 5$ and a Janus particle with $\beta_p = -10$.
Panels (e,g) show corresponding near-wall configurations, with the particle centered at $z_0/a = 2$; the squirmer is oriented normal to the wall, and the Janus particle is oriented at $\mathrm{rot}_x = \pi/3$.
Panels (b,f) report the surface slip velocity for squirmers with $\beta_s =\{-5, 0, 5\}$, and panels (d,h) report slip velocities for Janus particles with $\beta_p = \{-10, 1, 10\}$.
In all cases, open symbols denote BEM results and filled symbols denote PINN predictions.
For the squirmer, solid curves in panels (b,f) show the analytical slip distribution used as reference.
Velocities are normalized by the swimming speed $|\bm{U}|$.
}}
\label{fig:figure_2} 
\end{figure*}

\textit{Active Colloid Hydrodynamics.}---To model the dynamics of an active colloid, we consider a particle of radius $a$ moving in a low Reynolds number fluid described by the Stokes equations
\begin{equation}
\label{eq:1}
\eta \nabla^2 \bm{u} - \nabla p = 0 \, , \quad
\nabla \cdot \bm{u} = 0 \, ,
\end{equation}
where $\eta$, $\bm{u}$, and $p$ denote the dynamic viscosity, fluid velocity, and pressure, respectively.
For a free particle, the boundary condition at infinity is $\bm{u} \to 0$, and for confining boundaries, we impose no slip at the wall.
At the particle surface, the fluid velocity satisfies
\begin{equation}
\label{eq:2}
\bm{u} = \bm{U} + \bm{\Omega} \times (\bm{r} - \bm{r}_0) + \bm{u}_s ,
\end{equation}
where $\bm{U}$ and $\bm{\Omega}$ are the translation and angular velocities of the particle, which can be probed experimentally, and $\bm{u}_{s}$ is the slip velocity.
The particle satisfies the force- and torque-free conditions
\begin{equation}
\bm{F} = \int_{S} \bm{\sigma}\cdot\bm{\hat{n}}\, dS = 0 \, , \quad
\bm{\tau} = \int_{S} (\bm{r}-\bm{r}_0)\times(\bm{\sigma}\cdot\bm{\hat{n}})\, dS = 0 \, .
\label{eq:3}
\end{equation}
Here, $\bm{\sigma} = -p \bm{I} + \eta(\nabla \bm{u} + \nabla \bm{u}^T)$ is the hydrodynamic stress tensor, $\bm{\hat{n}}$ the surface normal, and integration is taken over the entire particle surface $S$.
These constraints, together with the Lorentz reciprocal theorem, yield the translational and angular velocities, $\bm{U}$ and $\bm{\Omega}$, for a given slip velocity distribution $\bm{u}_s$~\cite{lauga2009hydrodynamics}.

\textit{Squirmer Model.}---A standard model for the slip velocity of an active colloid is the squirmer model, in which
\begin{equation}
    \label{eq:4}
    \bm{u}_{s} = B_1 \sin\theta + B_2 \sin\theta \cos\theta ,
\end{equation}
where $\theta$ is the polar angle relative to the swimming direction $\bm{\hat{d}}$, and $
|\bm{U}| = 2|B_1|/3$ is the swimming speed.
In this work, we fix $B_1 =1$.
The dimensionless squirmer parameter $\beta_s = B_2/|B_1|$ classifies the flow field generated by the particle as a puller ($\beta_s > 0$), pusher ($\beta_s < 0$), or neutral swimmer ($\beta_s = 0$).
The Stokes equations for a squirmer in an unbounded domain admit an analytical solution, which we use as a reference~\cite{pak2014generalized,schmitt2016marangoni}.
We also employ BEM to compute flow fields, as it is an accurate, established method for Stokes flows~\cite{pozrikidis2002practical,bayati2019dynamics}.

\textit{Phoretic Janus Particle Model.}---For a phoretic Janus particle, we consider a generic model in which one hemisphere emits solute while the other hemisphere consumes it~\cite{moran2017phoretic,bayati2016dynamics}. 
This model describes a broad class of chemically driven active colloids, including enzyme-coated particles, thermoresponsive microgels, and bimetallic colloids~\cite{maiti2019self,schattling2015enhanced,ma2015enzyme,seo2013one,mou2014autonomous,paxton2004catalytic,dong2016highly}.  
The slip velocity resulting from solute gradients along the surface of the particle is
\begin{equation}
\label{eq:5}
    \bm{u}_{s} = -\, b(\bm{\hat{n}})\,\nabla_{s} C\big|_{s} \, ,
\end{equation}
where $C$ is the solute concentration, $\nabla_{s} = (\mathbb{I}-\bm{\hat{n}}\bm{\hat{n}})\cdot\nabla$ is the surface-gradient operator.  
The phoretic mobility $b(\bm{\hat{n}})$ depends on particle--solute interactions and may be positive or negative~\cite{anderson1989colloid,Derjaguin1993-ph}.  
We treat $b(\bm{\hat{n}})$ as constant on each hemisphere, denoted by $b_e$ and $b_a$, and define the mobility ratio $\beta_p = b_a/b_e$.  
We assume a purely diffusive solute so that the concentration field obeys the continuity equation
$\partial_t C = D \nabla^2 C$, where $D$ is the solute diffusivity.  
In our calculations, we assume solute relaxation is fast compared to particle motion, and the concentration field is treated as quasisteady, and satisfies the Laplace equation
$D \nabla^2 (C - C_\infty) = 0$, with $C_\infty$ the far-field concentration.
We provide additional model details in the End Matter for phoretic Janus particles.

\textit{PINN Framework.}---Figure~\ref{fig:figure_1}(b) gives an overview of the PINN framework used to infer the near-surface flow field and interfacial slip distribution from partial velocity data. 
A fully connected neural network with trainable weights and biases $(W,b)$ learns a mapping from spatial coordinates $\bm r = (x, y, z)$ to the hydrodynamic fields  $(\bm{u}_{nn},p_{nn}) = (\bm{u}(\bm r; W,b), \,  p(\bm r; W,b))$. 
These weights and biases are optimized by minimizing a composite loss function that couples physical constraints with available data,
\begin{equation}
{\cal L} = \lambda_1 {\cal L}_{\text{PDE}}+ \lambda_2 {\cal L}_{\text{BC}}+ \lambda_3 {\cal L}_{\text{FT}}+ \lambda_4 {\cal L}_{\text{wall}}+ \lambda_5 {\cal L}_{\text{data}} 
\label{eq:6}
\end{equation}
where the hyperparameters $\lambda_i$ control their relative weights.
The loss terms enforce the incompressible Stokes equations, boundary conditions at particle surface, force- and torque-free constraints on the particle, and no-slip conditions on walls.
Finally, ${\cal L}_\text{data}$ penalizes deviations between predicted and measured velocities at selected points.
The network is implemented in \texttt{PyTorch} and trained with the \texttt{Adam} optimizer and an adaptive learning rate until the total loss saturates below a prescribed tolerance.
Details of the architecture, loss terms, and hyperparameters are provided in the End Matter.

\textit{PINN Results.}---We first assess the PINN’s ability to recover hydrodynamic fields in an unbounded domain. 
Figure~\ref{fig:figure_2}(a,c) compares predictions of the flow field for a puller-type squirmer ($\beta_s = 5$) and a phoretic Janus particle ($\beta_p = -10$), respectively.
The PINN streamlines (blue) closely match BEM solutions (pink), including near-surface regions ($r/a < 1.2$) where no training data were supplied. 
The PINN also accurately predicts the surface slip velocity: Fig.~\ref{fig:figure_2}(b,d) compares the PINN, BEM, and analytical solutions for squirmers with $\beta_s = \{-5, 0, 5\}$~\cite{pak2014generalized,schmitt2016marangoni} and Janus particles with $\beta_p = \{-10, 1, 10\}$, showing excellent agreement in both magnitude and angular dependence of the slip.

To quantify the agreement between the PINN predictions and reference solutions, we compute the relative $L_2$ error between the PINN-predicted and analytical (or BEM) velocity fields,
$
E_{L_2}
= \lVert \boldsymbol{u}_\mathrm{PINN} - \boldsymbol{u}_\mathrm{ref} \rVert_2 \, / \,
       {\lVert \boldsymbol{u}_\mathrm{ref} \rVert_2},
$
where $\lVert \cdot \rVert_2$ denotes the Euclidean norm evaluated over all grid points.
Table~\ref{table:error_combined} reports the errors for squirmers and Janus particles across several values of $\beta$.
In all cases, the PINN achieves small relative errors in both the flow field and the slip velocity, comparable to those obtained from a BEM calculation.
For reference, the BEM solution for the squirmer model in an unbounded domain closely matches the analytical solution, with average $L_2$ errors of $0.018$ for the flow field and $0.057$ for the slip velocity.
These results show that the PINN maintains high accuracy across particle types and activity strengths and can recover the near-surface slip even without training data in this region.

To further assess the robustness and generalization capability of the PINN framework, we extend the analysis to particles near a no-slip rigid boundary.
In this setting, the network includes an additional loss term enforcing zero velocity at the wall. 
The presence of the boundary introduces strong hydrodynamic interactions that substantially alter both the flow field and the particle’s motion, providing a stringent test of the method.
Wall-induced coupling modifies translational and rotational velocities in a manner that depends on propulsion mechanism, distance from the wall, orientation, and the value of $\beta$.
In addition, for phoretic Janus particles, proximity to the wall perturbs the solute concentration field, thereby changing the slip distribution when the particle approaches the boundary. 
In contrast, for squirmers, the wall affects only the hydrodynamic flow field, which can alter the translational and angular velocity of the particle, but the slip velocity remains prescribed by Eq.~(\ref{eq:4}).

\begin{table}[t!]
\centering
\small
\setlength{\tabcolsep}{6pt}
\renewcommand{\arraystretch}{1.15}

\begin{tabular*}{\columnwidth}{@{\extracolsep{\fill}} ccccc}
\toprule
\textbf{Squirmer} & \multicolumn{2}{c}{\textbf{Unbounded}} & \multicolumn{2}{c}{\textbf{Near-wall}} \\
\cmidrule(lr){2-3}\cmidrule(lr){4-5}
$\beta_s$ & Flow $E_{L_2}$ & Slip $E_{L_2}$ & Flow $E_{L_2}$ & Slip $E_{L_2}$ \\
\midrule
-5 & 0.020 & 0.029 & 0.024 & 0.075 \\
0  & 0.020 & 0.007 & 0.073 & 0.055 \\
5  & 0.020 & 0.031 & 0.032 & 0.080 \\
\midrule
\textbf{Janus} & \multicolumn{2}{c}{\textbf{Unbounded}} & \multicolumn{2}{c}{\textbf{Near-wall}} \\
\cmidrule(lr){2-3}\cmidrule(lr){4-5}
$\beta_p$ & Flow $E_{L_2}$ & Slip $E_{L_2}$ & Flow $E_{L_2}$ & Slip $E_{L_2}$ \\
\midrule
-10 & 0.093 & 0.130 & 0.067 & 0.092 \\
1    & 0.042 & 0.044 & 0.059 & 0.046 \\
% 2    & 0.035 & 0.033 & 0.063 & 0.048    \\
10   & 0.053 & 0.046 & 0.055 & 0.063 \\
\bottomrule
\end{tabular*}

\caption{
Relative $L_2$ errors for flow and slip predictions for squirmer and Janus particles in unbounded and near-wall geometries. 
Unbounded squirmer errors use analytical solutions as a reference for both flow and slip; near-wall flow errors use BEM as a reference, while slip errors remain referenced to the analytical profiles. 
For Janus particles, all errors are computed relative to BEM solutions.
}

\label{table:error_combined}
\end{table}

Despite these complexities, the PINN accurately captures both the qualitative structure and quantitative details of the near-wall hydrodynamics, showing excellent agreement with the BEM solutions. 
Figure~\ref{fig:figure_2}(e–h) presents representative predictions for a puller squirmer and a phoretic Janus particle positioned above a wall. 
In both cases, the PINN reconstructs the wall-modified streamlines with high fidelity, including the enhanced circulation and geometric asymmetry generated by the boundary [Fig.~\ref{fig:figure_2}(e,g)]. 
The corresponding slip velocities [Fig.~\ref{fig:figure_2}(f,h)] also agree closely with BEM, with the PINN reproducing the angular dependence across different squirmer modes $\beta_s = \{5,0,-5\}$ and phoretic mobility ratios $\beta_p = \{-10, 1,10\}$. 
Although no analytical solutions exist in the confined geometry, quantitative comparisons with BEM show that the relative $L_2$ errors for both the flow and slip remain comparable to the unbounded-domain values (Table~\ref{table:error_combined}).
This consistency confirms that the PINN continues to infer near-surface slip and the full velocity field with high accuracy, even under strong hydrodynamic confinement.

\textit{Discussion.}---This work introduces a physics-informed neural network framework for resolving the interfacial mechanisms that generate motion in active colloids. 
By coupling sparse flow-field measurements with the Stokes equations and appropriate boundary conditions, the method reconstructs near-surface slip velocities and the full velocity and pressure fields with high accuracy. 
Its performance for both squirmers and phoretic Janus particles, in unbounded domains and near confining boundaries, shows that bulk hydrodynamic observations contain sufficient information to infer the slip velocity responsible for propulsion.

The ability to infer boundary conditions directly from partial observations suggests a broader strategy for characterizing soft and active matter~\cite{cichos2020machine}. 
Incorporating additional physical fields, such as solute concentration or temperature, would enable inference of phoretic mobilities, interfacial stresses, and other transport parameters. 
Extending the framework to interacting colloids or active droplets would allow data-driven reconstruction of collective hydrodynamic interactions~\cite{colen2021machine,hayano2025learning}. 
Integrating experimental flow data, where noise and tracer artifacts are unavoidable, represents an important direction, since PINN-based regularization can extract reliable physical information under non-ideal measurement conditions.

Although developed here for active colloids, the approach is general. 
It provides a practical strategy for inverse problems in complex fluids where boundary conditions, rheological parameters, or interfacial stresses are not directly accessible. 
Potential applications include microfluidic design~\cite{shrestha2025self,Merdasi2023-ao}, parameter and stress inference in non-Newtonian fluids~\cite{thakur2024viscoelasticnet,Reyes2021-oh}, near-interface particle dynamics~\cite{zhou2025drag}, and the stability and transport behavior of multiphase and biological interfaces~\cite{Nakamura2022-vw,movahhedi2023predicting,fu2025physics}. 
By linking macroscopic flow observations to microscopic interfacial physics, the framework provides a quantitative path forward in systems where traditional computational or experimental tools are limited.

A remaining challenge, shared with many PINN-based approaches, is handling sharp gradients or discontinuities across interfaces. 
Standard neural networks favor smooth representations, which can reduce accuracy when the true solution has rapid spatial variation. 
In our case, this manifests as diminished slip-velocity accuracy for a Janus particle very close to a wall ($z_0/a < 1.4$), where the slip profile exhibits steep variations. 
In contrast, the framework performs well for squirmers at comparable separations ($z_0/a > 1.1$). 
Recent advances including domain decomposition~\cite{he2022mesh,wu2022inn}, specialized architectures~\cite{li2025two,Qiu2022-jb}, and interface-aware activation functions~\cite{sarma2024interface} offer potential remedies. 
As these developments mature, they are likely to extend the range and robustness of physics-informed inference for complex interfacial flows.

%\bibliography{references}


\clearpage

\section{End Matter}  

\textit{Phoretic Janus Particle: Concentration Field}---For the phoretic Janus particle, we consider a particle with Janus balance $\chi = 0$, corresponding to a half-coated configuration. Mass conservation requires the emission and absorption fluxes $Q_e$ and $Q_a$ to satisfy $S_e Q_e - S_a Q_a = 0$, where $S_e$ and $S_a$ denote the areas of the emitting and absorbing hemispheres. At steady state this implies
$Q_a/Q_e = (1+\chi)/(1-\chi)$. The solute boundary condition on the particle surface is
\begin{equation*}
    -D\,\bm{\hat{n}} \cdot \nabla C(\bm{r}) =
    H(\bm{\hat{n}}\cdot\bm{\hat{d}})\,Q_e
    + \big[1 - H(\bm{\hat{n}}\cdot\bm{\hat{d}})\big]\,Q_a \, ,
    \label{eq:janus-bc}
\end{equation*}
where $\bm{\hat{n}}$ is the outward normal, $\bm{\hat{d}}$ is the Janus orientation, and $H(x)$ is a smooth indicator function. The far-field condition is $C(\bm{r}) \to C_\infty$ as $|\bm{r}| \to \infty$; in wall-bounded domains, $C$ satisfies a no-flux boundary condition at the wall.
To avoid discontinuities at the emitting--absorbing interface, we smooth the phoretic mobility profile as
\[
b(\bm{\hat{n}})/b_e =
    H(\bm{\hat{n}}\cdot\bm{\hat{d}})
    + \big[1 - H(\bm{\hat{n}}\cdot\bm{\hat{d}})\big]\beta_p ,
\]
with $b_e$ the mobility on the emitting side and $\beta_p = b_a/b_e$ the mobility ratio. We adopt a sigmoid transition,
\[
H(x) = \tfrac12\big[1 + \tanh(3(x-\chi))\big],
\]
to regularize the hemispherical discontinuity.
In unbounded geometries, analytical solutions of the Laplace equation for $C$ are available~\cite{bayati2019dynamics,bayati2024orbits}. For the general wall-bounded configurations considered here, we compute the solute field using BEM.

\textit{BEM Calculation Details}---Boundary element method (BEM) simulations are used to compute the solute concentration, fluid velocity fields, and particle kinematics. 
We solve the Laplace and Stokes equations in three dimensions using their boundary‐integral formulations, implemented with BEMLIB routines in \textsc{Fortran}~\cite{pozrikidis2002practical} and extended to handle active Janus particles~\cite{uspal2015self,bayati2016dynamics}. 
These calculations provide the reference solutions used for validation in both unbounded and wall-bounded geometries.

\textit{PINN Implementation Details}---The Physics-Informed Neural Network (PINN) framework was implemented in \texttt{PyTorch}~2.2.1. 
Separate neural networks were assigned to each output variable to improve convergence and stability. 
For rigid particles, the computational domain was divided into two regions---the particle surface and the exterior fluid---each represented by independent networks predicting the velocity components and pressure, yielding a total of eight networks. 
This modular design ensures continuity across interfaces and facilitates extensions to more complex geometries and boundary conditions.
All spatial derivatives required to evaluate the Stokes residuals and stress tensors were computed via automatic differentiation.

The network architecture consisted of nine fully connected hidden layers with 128 neurons per layer. 
Weights were initialized using PyTorch's default Kaiming-uniform initialization, and the \texttt{Swish} activation function $f(x)=x\,\mathrm{sigmoid}(x)$ was used throughout to ensure smooth second derivatives required by the governing equations. 
Alternative activations such as hyperbolic tangent or sine produced comparable results, whereas \texttt{ReLU} was avoided because of its vanishing second derivative.

Network parameters were optimized using the \texttt{Adam} optimizer with an initial learning rate of $5\times10^{-4}$ and an exponential decay (factor 0.1 every 1200 epochs).  
Training typically converged after approximately 4000 epochs, when the total loss ${\cal L}$ plateaued below $7\times10^{-5}$. 
The loss hyperparameters were
$\lambda_1=\lambda_2=\lambda_3=\lambda_4=1.0$ and $\lambda_5=5.0$. 
All hyperparameters---network depth/width, activation, and loss weights---were fixed based on preliminary experiments and standard PINN practices for fluid-dynamics problems~\cite{wang2023expert}.  
Batch size had minimal impact on performance; we used a batch size of 256.  
All losses were computed using the mean-squared-error norm.

Collocation points for the PDE residual were uniformly sampled within a cubic domain enclosing the particle. 
At each training step, a random subset of $N_\text{PDE}\approx5500$ points was used to evaluate ${\cal L}_\text{PDE}$, while $N_\text{data}\approx4300$ data points contributed to ${\cal L}_\text{data}$, and $N_\text{BC}\approx3721$ boundary points enforced ${\cal L}_\text{BC}$ on the particle surface. 
An additional $N_\text{test}\approx29791$ points were used to evaluate predictive accuracy. 
Velocity data near the particle surface ($r/a<1.2$) were excluded to emulate experimental limitations. 
All quantities were nondimensionalized using the particle radius $a$ and characteristic velocity $|\bm{U}|$ to improve numerical conditioning.

The computational complexity of a forward--backward pass through a single network is 
$\mathcal{O}(L N^{2})$, where $L$ is the number of layers and $N$ the number of neurons per layer, corresponding to  
$9\times128^{2}\approx 1.5\times10^{5}$ operations per training point.  
Because the PDE residuals require first- and second-order derivatives via automatic differentiation, the effective cost increases by a constant factor.  
The cost per epoch, therefore, scales as $\mathcal{O}\!\left(8\, N_\text{total}\, L N^{2}\right)$,
where $N_\text{total}$ is the total number of PDE, data, and boundary points.  
With the parameters used here and the available GPU resources, each training epoch required approximately $4$ seconds.

\begin{algorithm}[H]
\caption{Training a PINN for active particle velocity and pressure fields}
\begin{algorithmic}[1]
\State \textbf{Require:}
\begin{itemize}
  \item Eight neural networks, each predicting one output: exterior flow
  $u_x^{\text{out}}, u_y^{\text{out}}, u_z^{\text{out}}, p^{\text{out}}$ 
  and surface/interface flow 
  $u_x^{\text{s}}, u_y^{\text{s}}, u_z^{\text{s}}, p^{\text{s}}$.
  \item Training data: velocity values at training points.
  \item Collocation points for PDE residual evaluation.
  \item Boundary condition points.
  \item Learning rate, batch size, and number of epochs.
  \item Loss weights $\lambda_i$.
\end{itemize}

\State \textbf{Pre-processing:} prepare training data and batch collocation points.
\State \textbf{Create neural networks} with 9 hidden layers and $128$ neurons per layer.
\State \textbf{Initialize optimizer and learning-rate scheduler}.

\For{$epoch = 1, \dots, epoch_{\text{total}}$}
    \State Initialize losses: $\text{loss} = 0$.
    \For{each batch $(x, y, z)$ from PDE points}
        \State Zero gradients of network parameters $\bm{W}$ and $\bm{b}$.
        \State Predict velocity and pressure fields.
        \State Compute total loss: $\text{loss} = {\cal L} =\smash{\sum_i}\lambda_i\,{\cal L}_i$.
        \State Compute gradients: \texttt{loss.backward()}.
        \State Update $\bm{W}$ and $\bm{b}$ using \texttt{optimizer.step()}.
        \State Accumulate batch losses.
    \EndFor
    \State \textbf{End For}
    \State Record total loss for this epoch.
    \State Update the learning rate using the scheduler.
    \State Save model weights and loss values for diagnostics.
\EndFor
\State \textbf{End For}
\State \textbf{Save final trained parameters $\bm{W}$ and $\bm{b}$}
\State \textbf{Prediction:} Apply the trained model on test data points to predict the velocity.

\end{algorithmic}
\end{algorithm}

\end{document}